# Neural association between musical features and shared emotional perception while movie-watching: fMRI study

Leonardo Muller-Rodriguez

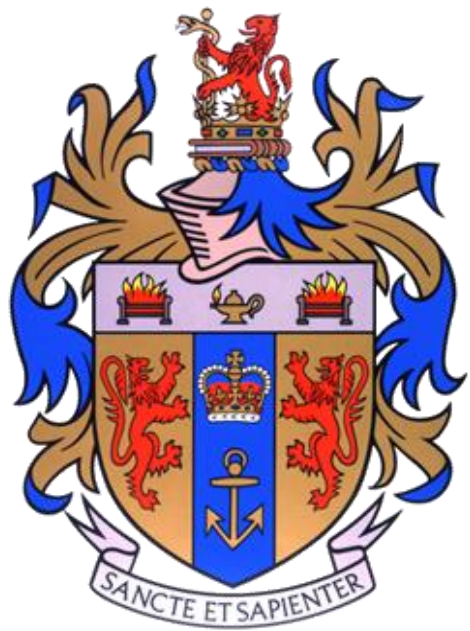

Supervisor: Dr. Owen O'Daly

The department and university affiliation in which the research was carried out
e.g Department of Basic and Clinical Neuroscience
Institute of Psychiatry, Psychology & Neuroscience
King's College London
University of London

**Thesis in partial fulfilment for the degree of MSc in Neuroscience September, 2021.**

**Word Count**

| Word Counts: | *Title*: | 14 |
| --- | --- | --- |
| | *Abstract* (max. 350): | 327 |
| | *Main Body* (7,500-10,000; word count does <u>not</u> include Title, Abstract, References and Appendices; see Module Handbook): | 9915 |

Personal Statement:

- A publicly available neuroimaging dataset by Hanke et al 2016 was chosen and analysed by Leonardo Muller (OpenNeuro). Thus, experimental design and acquisition was performed by Hanke et al 2016.
- Hypothesis and methodology were determined by Leonardo Muller and advised by Dr Owen O'Daly
- Data extraction of features was performed in Python and in MATLAB by Leonardo Muller
- Data analysis was performed in MATLAB by Leonardo Muller
- Dr Owen O'Daly assisted creation and guided editing of MATLAB codes.
- Dr Owen O'Daly oversaw the preparation of this thesis report.

**ABSTRACT**

The growing use of naturalistic stimuli, such as feature films, makes research on emotions come closer to ecologically valid settings within brain scanner environments, e.g. functional magnetic resonance imaging (fMRI). Music is another cultural artefact known to induce emotions. Film soundtracks tend to be designed to support the narrative content to enhance the emotional experience of the audience. However, the neural underpinnings of consistent (shared) emotional perception during movie-watching, and its relation to the soundtrack remain unknown.

In this study German viewers (n=15) watched the (German dubbed) film 'Forrest Gump' in a (3T) fMRI scanner. The reported consistency of arousal, positive and negative valence of film events were correlated with spectrogram, and tempo information extracted from the soundtrack. For the neuroimaging data, four regions of interests (ROI) related to audio-visual input, and narrative content were explored: superior temporal sulcus, amygdala, inferior frontal gyrus and precuneus.

Pearson's correlation revealed strong link of spectral and tempo acoustic features with emotional features (Bonferroni corrected). We also compared BOLD response associated with consistent (shared) positive, negative valence and high arousal film moments to film events with those characterised by low agreement regarding emotional content in our ROIs and found significance in superior temporal sulcus (STS) and precuneus ($p_{FWE}<0.05$ peak level).

No significant responses were seen in the amygdala or inferior frontal gyrus ROIs, but distributed brain activity observed in an exploratory whole-brain analysis ($p_{FWE}<0.05$ cluster extent) reveals frontal region activity linked to negative valence, as well as the obvious occipito-temporal lobes in response to the audio-visual naturalistic film.

Superior temporal sulcus and the precuneus were active during shared perception of emotionally valenced and high arousal events. Particularly, we showed that the precuneus encodes valence from low and high frequency auditory information. Nevertheless, naturalistic stimuli need to be designed for experimental comparison to improve interpretation of results and accompanied by stronger measurements of emotionality and shared cognitive processes among viewers. Overall, naturalistic stimuli remain a promising avenue as diagnostic and therapeutic tools for emotion regulation deficits.

# Table of Contents

# INTRODUCTION

Humans have evolved to process salient audio-visual information from their surrounding ecological environment. With the advent of technologies and knowledge throughout history, cultural artefacts such as music and cinema were crafted to provide a means to experience ecologically valid music-induced emotions in elegantly decorated concert halls, theatres, and immersive cinema rooms - now commonly detached from ecological survival.

## A - Current state of human emotion's field

Emotions are a first person, subjective expression of an individual's relationship to its environment and interpersonal encounters (1), resulting in body states on which information is understood and decisions are made (2). Remarkably, some researchers divide human emotions into two classes: basic and non-basic – mainly distinguished as non-social and social emotion labels, respectively (3).

Basic emotions are identifiable universally, in all cultures (fear, happiness, sadness, disgust, anger). Studies by Ekman in different cultures first demonstrated their universality by using grayscale static facial expression of actors displaying the mentioned basic emotions (1). However, the temporal evolution of emotions, such as facial expression and vocalisation fluctuation, provide emotional information that may be categorised differently depending on the cultural upbringing of individuals and influence of environmental cues, e.g. music. Interestingly, it is very common to confuse some basic emotions for another, such as fear with surprise and anger with disgust. This is because the prototypical facial muscle movements that discriminate emotional expression, aka action units, are shared at different time points between the temporal evolutions of basic emotions (4).

Emotion categorisation is supposed to function like word recognition. While receiving dynamic linguistic information, the individual narrows down options of potential word and sentence meanings until one isolation point: i.e., a certain word is reached, and no other word prediction seems feasible. Therefore, we recognize basic emotions by narrowing down to the emotion category that best matches the facial expression, gestures, posture/gait, vocalizations and semantic information experienced (4). Understanding metaphorical gestures accompanying speech is shown to occur in the inferior frontal gyrus (IFG) (5).

### 1 - Neural distribution of emotions

Emotion processing uses a distributed set of interacting and overlapping neural networks and loops which activate to a greater or lesser extent depending on the environmental (6) and mental (imagery) context (7). These networks involve common emotion brain regional hubs (particularly well-connected regions) in unique configurations per subjective experience of emotions (3, 8). The amygdala, the anterior cingulate cortex (ACC), medial pre-frontal cortex (mPFC) and posterior superior temporal sulcus (pSTS) are amongst the emotion hubs characterised during salient audio-visual sensory processing (9, 10). These hubs are at the intersection of the diverse neuronal circuitry (6). Moreover, emotion integration regions, such as the default-mode network (DMN), are the last to be active in the cascade of functional activity (11).

Importantly, emotion-related hub co-activations invoke several networks simultaneously, to a degree depending on spatiotemporal conditions. The signal does not transmit via a pre-determined number of neurons, but rather continually modulates pathways via neuroplasticity. In addition, theories of one-way controlled (top-down or bottom-up) flow narrow the understanding of existing reciprocal signalling - whether direct, such as amygdala to mPFC and vice versa, or indirect, such as amygdala to thalamus to mPFC. Thus, the brain is heterarchical rather than hierarchical (6).

## 2 – Emotion markers

Curiously, emotional expression dynamics tend to be processed with greater facility and intensity as the number of congruent modalities of presentation are increased (4, 12-16) such as audio-visual versus visual stimuli. This accords with findings from investigations of sensory illusions arising from the integration of audio-visual cues of varying degrees of congruency, such as the sound-induced flash illusion and the McGurk effect (17, 18).

It is argued that physiological processes, like heart and breathing rate, characterised by changes in affect (valence and arousal) are not labels of emotions, but rather are deciphered by the experiencer via emotional labels, like fear or excitement, depending on their memories of interpreting similar situations (19). In addition, brain architecture similarities and differences between animals denote species-general and species-specific computations, respectively, characterizing the degree of abstraction utilized for *ad hoc* concepts in emotional episodes (19). Therefore, defining certain ethological behaviours as types of emotions does not imply objective measurement of emotions but rather showcases the scientists' inference at play. Although this viewpoint is contested by some scientists who believe emotion categories can be objectively characterised by stimulus-reactivity, most research up-to-date tend to discuss emotion concepts (e.g. fear) rather than situations that evoke fear (19). Emotions research carried using static and simple, principally visual, stimuli with low ecological validity, question the generalizability of results for understanding brain functions which are widely spread and complex (20). Moreover, functional magnetic resonance imaging (fMRI) and laboratory settings, although isolated from everyday environments, are currently the only means to understand brain activity experimentally. Therefore using more ecologically valid stimuli approaches experimental conditions to daily life environments (21).

## 3 - Musical emotions

Importantly, emotion dysregulation is identified as a transdiagnostic factor in several clinical psychopathologies, impoverishing social life. Mirroring of other's bodily and facial expressions for inferring interpersonal emotions has shown to activate pSTS. Observing social rejection in others activates ACC and anterior insula, replicating their experience (10). Strikingly, training emotion regulation greatly improves patients' illness severity, by modulating emotional responses and reaching desirable goals (22, 23).

Music is an effective naturalistic stimulus known to induce emotions, hence the interest in music therapy to aid patients (24). Nevertheless, its ecological validity in respect to human evolution can be open to debate, arising during the Middle Palaeolithic period (25). Moreover, music possesses no clear ethological validity. There appears to be entrainment – synchronising to the musical beat – in several species such as parrots, sea lions and elephants, but the evidence is scarce for musical appreciation (26). Nevertheless, cultures around the world choose to listen to sounds and music to feel emotions (27). There are exceptions due to neurological damage, or healthy individuals void of musical pleasure (28), nonetheless music tends to induce at least 3 levels of pleasure: visceral due to physical sound vibrations, melodic and rhythmic memory, structural and cultural knowledge of musical pieces (27).

Music-induced emotions tend to be characterised in terms of their associated arousal (29) and valence. The main contributors of these measures are the fluctuation of musical features such as tempo speed (30) and timbre (consisting of temporal envelope modulation and fine structure - spectral distribution of the harmonic frequencies of sounds besides its overall loudness and fundamental pitch) (31, 32). Timbre is principally perceived via its fine structure, as demonstrated by lower capacity in cochlear implant users, who have reduced ability to detect fine structure (33). Subtle changes in timbre and volume, known as emotional expressivity of the musical instrumentalists, or greater modifications in timbre are encoded in the auditory cortices, according to (34-36). Linguistic prior knowledge of lyrics or music-genre label also seem to elicit emotional reactions independent of musical features, alluding to the influence of prior musical emotional experiences on listening. This priming permeates the subsequent musical experience, regardless of linguistic matching to the music genre (37).

## B – Naturalistic dynamic stimuli for emotions research

### 1 - Music in films

Generally, music used in movies are novel to the viewer because of its creation for that specific audio-visual narrative. Movies are, therefore, a great stimulus to study music-induced emotions due to the emulation of rich visual information present during music-listening activities and the director's drive for music and visual emotional congruency which provokes faster sensory processing for human brains (4, 12).

It is possible that film directors' and music composers' intricate audio-visual stimulus manipulation may underlie acquired intuitions about underlying brain architecture and functionality like basic audio-visual congruency. For example, congruent task-irrelevant modality facilitates task performance, whereas if it is incongruent, it disrupts performance (4). Moreover, in a simple visuo-spatial task, visual alerts disrupted to a greater extent the visual task than auditory alerts (38).

Furthermore, when emotionally salient stimuli, faces and music, are presented together, common in films, emotional valence congruency results in greater subjectively experienced emotional intensity. This emotional intensity is reduced when the stimuli are emotionally incongruent, or the visual stimulus is absent. Interestingly, attentional distribution theory states that positive emotions widen the breadth of attention whereas negative/unpleasant audio-visual emotions narrow it (14, 39, 40). It is suggested that positive music and face congruency broadened the participant's attention to the visual information, perceiving greater emotional facial information compared to negative music (14).

Interestingly, face and pitch discrimination seem to share integrational brain areas. Damage to the fusiform gyrus resulting in impaired facial recognition – known as prosopagnosia -, seems to show a link with congenital amusia, more specifically auditory pitch discrimination impairment, notably in developmental cases, i.e. acquired from birth. Neural network commonality is unclear due to small sample size studies, yet pSTS lesions, situated in proximity to the primary auditory cortex, are shown to cause facial expression impairment (41, 42). Moreover, greater neural activity over time was measured with single-unit electrode and fMRI in the (face patch) anterior fundus of the pSTS of macaque monkeys when exposed to audio-visual pan-threat vocalisation over audio or visual only (12). The pSTS has functional connectivity with the inferior frontal gyrus (IFG) – higher-level bimodal integration hub - part of the auditory and language dorsal-stream regions (43).

### 2 - Naturalistic narrative stimuli

Nevertheless, due to the ability of movies to strongly elicit emotions, participants exposed to naturalistic stimuli are therefore studied in a more ecologically valid context. The wide appeal of cinema to the public reflects an evolutionarily shared brain architecture and subsequently similar patterns of neural responses, as demonstrated by the effectiveness of inter-subject correlation (ISC) analysis of brain activity while movie-viewing. ISC is a recent neuroimaging analysis method that identifies synchronous and shared brain activity amongst participants in an fMRI experiment, whether similar decrease or increased neuronal activity in brain regions (44).

Most studies using movies have been of exploratory nature but have revealed the neuronal impact of movies. Films such as *The Good, the Bad and the Ugly (GBU)*¸ show higher ISC brain activity than uncut footage of everyday, dynamic, urban events. This was also reflected in eye gaze position variance which is highly correlated between humans in semantically richer and coherent movie stimuli such as *GBU* (45, 46). There is even a moderate inter-species correlation of eye gaze position between humans and monkeys (47).

It is important to note that low ISC may indicate variability caused by idiosyncratic intellectual focus, such as in art films with ambiguous image flow, also seen in the use of long uninterrupted, everyday shots. It may also imply higher daydreaming and attentional disengagement with the film. Highly controlled dynamic juxtaposition of images results in higher ISC (44). On the other hand, DMN activity is related to an inward attentional focus (45), or daydreaming of complex social situations (11) and is shown to be less active during high suspense (45). Strikingly, audio-book narratives, which tend to show high ISC in auditory cortex and low in visual cortex – silent films show the inverse (44) -, elicit the same inter-subject responses in DMN for Russian listeners and English listeners for the same spoken narrative in their respective languages. Furthermore, DMN activity, here, was coupled with the narrative content (48). Recalling semantic information in short temporal receptive windows, such as associating a verb to its subject in a sentence, is shown to activate sensory cortices, **whereas** longer temporal receptive windows tend to activate DMN (e.g. precuneus) and right pSTS, by remembering characters presented 10mins prior to the present scene. The latter shows slower ISC frequency, unsynchronized with sensory information, indicating the slower processing of plot formation (11).

Valence of narratives has shown to modulate social affiliation networks, such as increased connectivity between social aversive network and dlPFC and parietal cortex while listening to a patient's insecure-dismissive childhood parental relationship. This correlated with an individual's attachment anxiety (negative models of self, worry of rejection by others) and evaluation of friendliness (49). Interestingly, brain activity between a narrator telling a story and the listener is correlated by a delay, and the extent to which neural alignment corresponds between the speaker and listeners, the better the communication as assessed with post-scan tests (48).

Ventro-mPFC and DMN to hippocampus functional connectivity increases the more congruent the narrative context exposure is, such as understanding the narrative towards the final scenes of *Memento,* or exposure of a congruent contextual cue one day compared to minutes before viewing that film. In addition, mPFC activity is shown at moments when events in *Memento* are shed with new information (11).

Increased narrative suspense is shown to decrease peripheral vision as demonstrated with decreased anterior calcarine sulcus activity – which processes visual periphery – and suppressed processing of peripheral checkerboards, relative to moments of decreasing suspense. Interestingly, congruent music soundtracks was significantly related to enhance suspense, compared to incongruent or no music, but it is only high narrative suspense that increased memory recall of story events regardless of underlying soundtrack – a surprising finding which may reflect low sample size (45). High auditory ISC demonstrates soundtrack effectiveness, as seen with Alfred Hitchcock's *Bang! You're Dead* compared to *GBU* and everyday urban events (44).

## C – Aims of this study

Therefore, emotionally congruent music cues in movies support narrative communication, retention, and importantly, increase emotional impact for the audience. Fascinatingly, cinema, an audio-visual cultural artefact, is approximately a century old, yet is incredibly effective in eliciting emotions. While variation is likely intrinsic, recent work (studyforrest.org study (50)) reported the periods of high agreement and disagreement on the judgement about an actor's portrayed emotions by participants who viewed *Forrest Gump*. However, the cause of fluctuation of inter-objective agreement of emotion portrayal for this film remains unclear.

One possibility is that high congruency in perceived emotions (portrayed by an actor) coincided with periods of emotionally congruent music (measured via certain musical features). While there is plenty of evidence for music modulating emotions and perceptual processes, the degree to which music can drive shared emotional activity in groups of humans, although anecdotally obvious, remains to be fully understood.

Hence, we posit three hypotheses:

1. We expect that musical timbre or tempo will correlate with the consistency regarding the arousal and valence of conveyed emotions.

2.      We expect that moments of high consistency in both perceived arousal and valence will be associated with activation of brain regions pSTS (role in mirroring other's emotions notably during audio-visual congruency (10, 12)), amygdala (important emotion-related role (9)), precuneus (the door to the DMN, important during storytelling (11)), and IFG (role in emotion categorisation, like matching semantic information to gestures (5), possibly useful for agreement on emotion-related labels, e.g. arousal and valence).

3.      We expect that in brain regions IFG, pSTS, precuneus and amygdala, trial-to-trial variation in activity during emotional valence processing will correlate with musical timbre and tempo. The specific musical features chosen for analysis will depend on the results from hypothesis 1 & 2.

# METHODS
*Note on methods*

This study analysed publicly available data from https://studyforrest.org. In this Methods section, I will clarify the work done by other researchers (studyforrest), and the work carried out by Dr Owen O'Daly and me. The neuroimaging and accompanying studyforrest dataset were downloaded from git using datalad commands within a conda environment. A 7 Tesla fMRI scanning of audio-description of Forrest Gump was also part of this larger study, however I will limit the description of the data and methodology used in the thesis, but if the reader wants to know more about the full study, it is recommended to explore the publications and public datasets found in the above website.

## Participants
*Carried out by studyforrest*

   We used a publicly available neuroimaging dataset from studyforrest.org exploring naturalistic stimuli focused on the audio-visual movie Forrest Gump. In this experiment, fifteen right-handed participants (mean age 29.4 years, range 21–39, 6 females, native language: German) (51) viewed an audio-visual, German dubbed version of Forrest Gump, the researcher's cut, i.e. a shortened version of the film which omitted irrelevant scenes to the narrative to reach 2 hours. The same participants had previously participated in an audio-description session of Forrest Gump in a 7T fMRI, but that dataset will not be analysed due to the great difference in MRI parameters used in both sessions.

## Stimulus
*Presented by studyforrest*

   We signed a written authorisation with the lead researcher of the studyforrest.org project (Dr M. Hanke) which states that the movie will only be used for the purpose of this thesis. During the experiment, the researcher's cut (number of volumes acquired per movie segment: 451, 441, 438, 488, 462, 439, 542, and 338, segments 1-8, see) of the Oscar-winning film 'Forrest Gump' (R. Zemeckis, Paramount Pictures, 1994, dubbed German soundtrack), high-resolution Blu-ray disk release (2011; EAN: 4010884250916) was presented for the participants to view (50). For the mirror projection of the visual stimulus and its dimensions please refer to the 'Stimulation setup' in the Methods section of Sengupta et al, 2016. For audio and visual re-encoding processing done to stimulus please refer to Stimulus in the methods section of Hanke et al, 2016. Summarise of relevant parameters, full methods are to be found in.

Participants were instructed to 'enjoy the movie' and to maximally inhibit any physical movement apart from eye movement. Prior viewing the movie, participants were given the credit's soundtrack to put volume at a desirable level. After each segment, participants were asked to rate (out of 4) how deeply they engaged with the story and were allowed a break from the scanner which lasted approximately 10mins amongst the group. Prior each segment, the eye tracker was calibrated, and gaze accuracy was validated.

| Part from Original | Start time | End time | Start frame | End frame | REMOVED SECTIONS | |
|---|---|---|---|---|---|---|
| 1 | 00:00:00 | 00:21:32 | 0 | 32312 | 00:21:32 | 00:24:13 |
| 2 | 00:24:13 | 00:38:31 | 36349 | 57798 | 00:38:31 | 00:38:58 |
| 3 | 00:38:58 | 00:57:19 | 58470 | 85997 | 00:57:19 | 00:59:31 |
| 4 | 00:59:31 | 01:18:14 | 89293 | 117351 | 01:18:14 | 01:20:24 |
| 5 | 01:20:24 | 01:34:18 | 120616 | 141457 | 01:34:18 | 01:37:14 |
| 6 | 01:37:14 | 01:41:30 | 145869 | 152269 | 01:41:30 | 01:42:49 |
| 7 | 01:42:49 | 02:09:51 | 154244 | 194792 | 02:09:51 | 02:22:00 |
| **Original length** | 00:00:00 | 02:22:00 | n/a | | n/a | |

**Table 1:** 7 Isolated parts from the original movie used in the study.
*The isolated scenes were considered crucial for the narrative, whereas the removed sections were deemed inessential to the plot. Timestamps relate to the original movie and are given in HH:MM:SS.FRAME format, and refer to the 2002 DVD release (PAL-video, 25 frames per second, DE103519SV).*

| Segments | RESEARCH CUT | | fMRI Volumes |
|---|---|---|---|
| 1 | 00:00:00 | 00:15:06 | 451 |
| 2 | 00:15:06 | 00:29:40 | 441 |
| 3 | 00:29:40 | 00:44:18 | 438 |
| 4 | 00:44:18 | 01:00:38 | 488 |
| 5 | 01:00:38 | 01:16:06 | 462 |
| 6 | 01:16:06 | 01:30:48 | 439 |
| 7 | 01:30:48 | 01:48:56 | 542 |
| 8 | 01:48:56 | 02:00:15 | 338 |

**Table 2:** Segments from the edited research cut, and its corresponding number of brain scans, timeline analysed in this study.
*The seven isolated parts were edited together and then split again into eight segments. fMRI scans were acquired while participants viewed each segment mentioned here. Timestamps relate to the edited (research cut) movie and are given in HH:MM:SS.FRAME format.*

### T2*-weighted functional images
*Carried out by studyforrest*

In this thesis, we analysed experimental data acquired via a whole-body 3 Tesla magnetic resonance imaging (MRI) Philips Achieva dStream, equipped with a 32-channel head coil. The following parameters were set for the acquisition of the T2*-weighted echo-planar images: 2 s repetition time (TR), 30 ms echo time, 90° flip angle, 1943 Hz/px bandwidth, parallel acquisition with sensitivity encoding (SENSE) reduction factor 2. Each volume contained 35 axial slices with an in-plane resolution of 80×80 voxels (voxel size: 3.0x3.0×3.0 mm), 240 mm field-of-view (FoV), anterior-to-posterior phase encoding direction) with a 10% inter-slice gap were recorded in ascending order—practically covering the whole brain (51).

**T1-weighted structural images**
*Carried out by studyforrest*

The structural T1-weighted images were acquired with a 3-Tesla Philips Achieva equipped with a 32-channel head coil using standard clinical acquisition protocols. Each subject's T1-weighted image contained 274 sagittal slices (FoV 191.8×256×256 mm, voxel size: 0.7 mm$^3$). A 384×384 in-plane reconstruction matrix (0.67 mm isotropic resolution) was recorded using a 3D turbo field echo (TFE) sequence (scan duration 12:49 min, flip angle 8 degrees, echo time 5.7 ms, TR 2.5 s, inversion time 900 ms, bandwidth 144.4 Hz/px, Sense reduction AP 1.2, RL 2.0).

**Musical features extraction**
*Performed out in present study*

Dr Hanke shared 8 audio-visual segments forming the German dubbed research cut of Forrest Gump as .mkv files, plus the additional credits scene which was visually empty (had a fixation dot on a grey screen) but contained all the soundtrack used while the participants adjusted the volume prior to scanning. For the audio analysis, the audio-visual .mkv files were converted to audio .wav files using moviepy, a Python library. Subsequently, each segment was divided into 4 second audio chunks, with a sliding window of 1 second to perform the spectrogram and tempogram analysis over 4 second sections for every second of the research cut's soundtrack. The wave file chunking was done using the pydub.AudioSegment and pydub.utils.make_chunks Python libraries. These audio chunks were then used to perform short-time Fourier Transforms (figure 1), also known as spectrogram analysis, over 4 second windows for every second, then concatenated back into the full length of the segment, divided into three equal frequency bands characterised as low, mid, and high frequencies. For the tempo, tempo information was acquired for every audio chunk per segment, similarly to the spectrogram, to assembled into the duration of the segment. The spectrogram and tempogram were performed and scripted in Python (Spyder 5, Python 3.9.5, packages: librosa.stft, librosa.feature.beat, numpy, matplotlib, os, itertools).

Once produced, the spectrogram allowed us to assess the timbral properties of the soundtrack for each segment of the movie (figure 2). We could therefore also assess the tempo fluctuation of the soundtrack throughout the research cut (figure 3).

The Fourier Transform (figure 1) allows us to transform sound signal in the time domain into frequency data for that signal. More specifically, we figure out the amplitude and phase of sinusoidal waves for each identified frequency between + ∞ and -∞ that form the source sound signal when summed together (52, 53).

$$X(f) = \int_{-\infty}^{\infty} x(t) \times e^{-i2\pi ft} \, dt$$

**Figure 1: Forward Fourier Transform.**

*We multiply the signal x(t) by the negative exponential values that correspond to rotations in cycles/second, as following Euler's formula (not shown here) which describes sinusoidal waves with real and complex components. X(f), with f in Hz, represents all the pure sinusoidal waves that correlate with the signal, each obtained by the mentioned multiplication. The integral sums up the magnitudes of each frequency that make up the signal (52).*

In this thesis, the Fourier Transform was performed in windows (or short snippets) of the soundtrack. This computation is repeated as the window slides through the soundtrack, (partly) overlapping Fourier Transforms on the previous window to compensate for the time/frequency resolution problem - either better time resolution (shorter window) but reduced data points for frequencies, or the inverse. This obtains all signal frequencies for all windows (54).

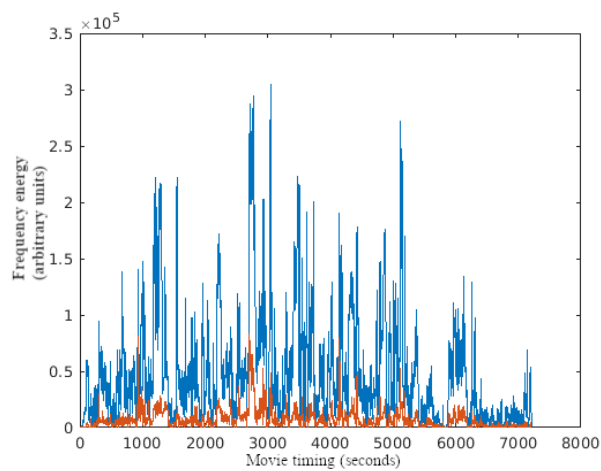

**Figure 2:** Energy of low frequencies (blue) and mid frequencies (red) throughout all 8 segments (x axis: time in seconds), displaying spectral energy fluctuation throughout the research cut.

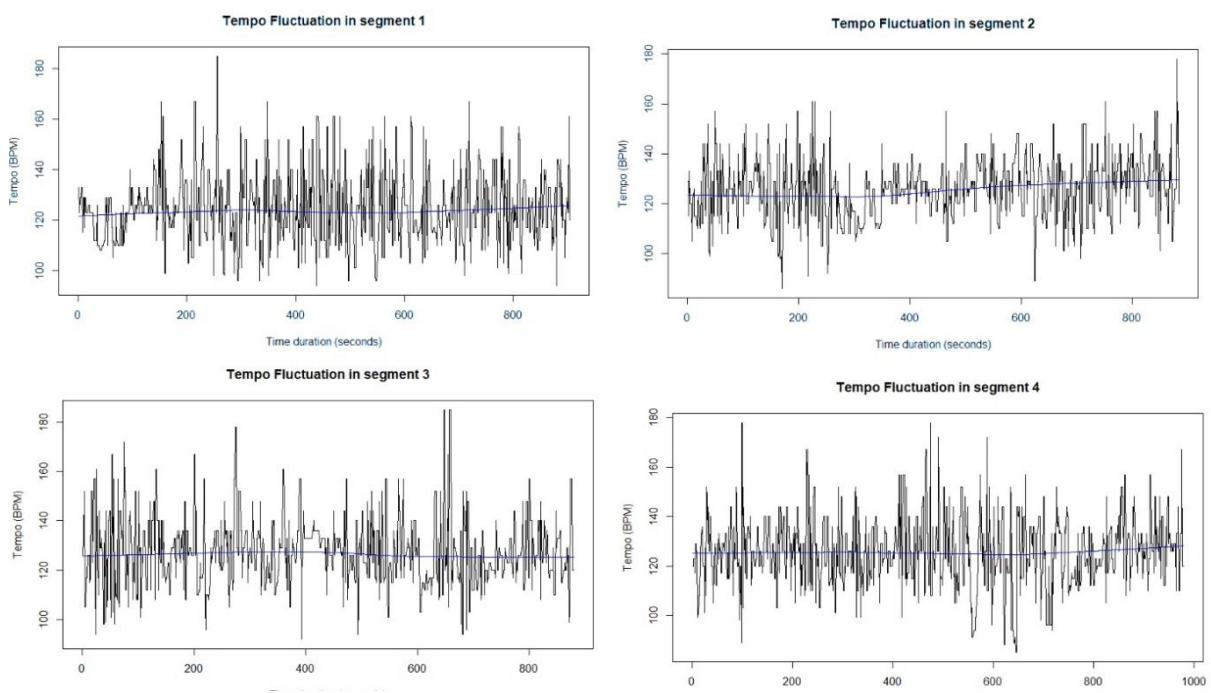

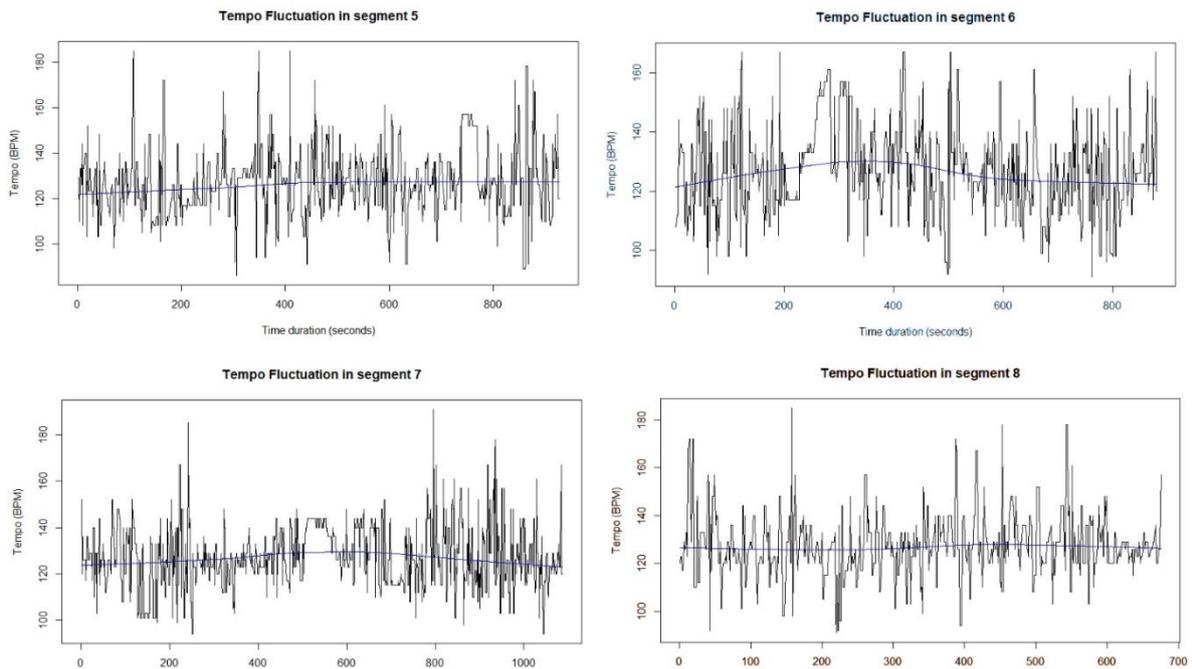

**Figure 3:** Tempo fluctuation (in BPM) over time (in seconds) of the soundtrack for each of the 8 movie segments. *The tempo mean stays around 120BPM throughout the whole film, suggesting that this is the ground tempo of the film. For all 8 segments, the tempo fluctuates between 100 to 140 BPM approximately, mainly in the fast tempo range (>120BPM) following (30).*

**Emotional onsets' extraction**
*Carried out in present study*

9 female students were asked to provide a binary value for valence (positive or negative) and its arousal (high or low) of emotional portrayal during randomized audio-visual movie events specified by Advene (https://www.advene.org/), an open source video annotation tool, a task that lasted approximately 30 hours per observer. Other annotations were conducted by the research team, but not used in this study. The randomization removed slow mood effects of the narrative, and the scenes of emotional portrayal were delimited by locational changes. Data quality was controlled by an algorithm that verified that all annotation entries were completed (55). The observer's account of the indicator variables of portrayed emotions were then averaged to give the inter-objective agreement on these emotional features during the film (55). We then extracted these onsets from the dataset using Matlab code (see Lab Book). In the absence of prior published literature to guide our choice, we arbitrarily considered emotion-perception trials to have high inter-objective agreement if their score had 50% or greater agreement across participants (where only high arousal agreement was considered).

**Pearson's correlation between emotional and musical features**
*Analysed in present study*

For the correlation of emotional ratings and musical features (hypothesis 1), the emotional values were taken directly from the studyforrest dataset, without modification. The timing information for the relevant events or features were used to construct a timeseries representing that features representation during each segment. These segments were combined and then the correlation between each musical feature and the chosen emotional trial types was tested.

We created a correlation matrix by obtaining Pearson's correlation coefficient for each combination concerning the 5 musical features (tempo in BPM, volume in RMS, low, mid, and high frequency band energy) and 3 emotional features – the timecourses of inter-objective agreement for arousal, positive and negative valence. Note that here every emotional onset was included with an amplitude equal to level of agreement among participants. This was done in Matlab R2018b (https://uk.mathworks.com/), using the Matlab function for Pearson's correlation coefficient: corrcoef() (see Lab Book for all mentioned scripts).

**Functional neuroimaging data pre-processing**
*Carried out in present study*

Data was preprocessed using Statistical Parametric Mapping 12 (SPM12 v7487; www.fil.ion.ucl.ac.uk/spm/) commands within IBM Wisconsin Matlab R2018b (https://uk.mathworks.com/) script. Preprocessing consisted of seven steps. The images were first *reoriented* towards the anterior commissure, both for T1-weighted and functional MRI. This was followed by *unified segmentation* of the T1-weighted images into grey, white matter, cerebral spinal fluid, a process which also generates a deformation field that can be used to warp data from native space to standard (MNI) space during normalisation. Subsequently, the *slice-timing* effects of ascending slice acquisition of functional images were corrected. The images were then *realigned* using rigid-body registration using 6 degrees of freedom as affine transformations on the 3 axis (3 translations and 3 rotations) to correct for volume-to-volume head motion. Motion correction did not exceed the voxel size (i.e. 3mm) in any case. We then *co-registered* functional images onto the same subject's T1 structural scan (similarly reoriented) used 12 degrees of freedom (6 rigid-body registration + 3 shears and 3 zooms). The data was *normalised* using linear 12 degrees of freedom affine registration followed by high-dimensional non-linear deformation, to warp the brains onto a common reference space (Montreal Neuroimaging Institute, International Consortium for Brain Mapping, (MNI) template space). Finally, a *3D smoothing* kernel of 6mm full-width at half the maximum height (FWMH) was applied to the normalised functional images to increase signal-to-noise ratio.

## Region of interest (ROI) masks
*Carried out in present study*

We used the AAL3 atlas from SPM12-compatible WFU_PickAtlas toolbox to generate four hypothesis-driven bilateral ROI masks (see figure 4): superior temporal sulcus (STS), precuneus, amygdala & inferior frontal gyrus (pars triangularis and pars operculum).

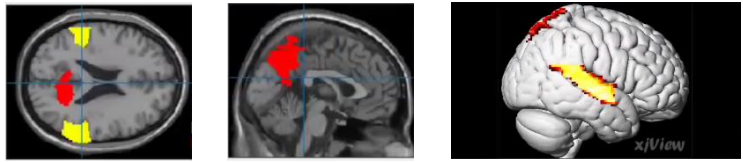

**Figure 4:** Hypothesis-led ROIs chosen a priori on the basis of past literature implicating these regions in processes of central interest to the project, namely audio-visual integration and mirroring of other's emotions, narrative encoding and daydreaming. *In red* the bilateral precuneus ROI mask. *In yellow*, the bilateral STS ROI mask. Images from xjview SPM12 toolbox.

We tested a smaller set of more focused hypotheses related to brain activation associated with musical features. As our correlation analysis failed to demonstrate an association between tempo and arousal in the non-imaging correlational analysis, we did not explore the relationship between tempo and the BOLD responses associated with trials where participants consistently reported high arousal (see Results), i.e. shared emotional perception. Furthermore, another reason to omit the tempo from the neuroimaging analysis is that it *primarily* displayed fluctuation within the fast range (>120BPM; figure 2). Previous neuroimaging study (30) investigating the role of tempo compared the fast range to mid (76-120BPM) and low (60-76BPM) ranges, which are required to give experimental weight for a neuroimaging study.

Due to an absence of significant emotion-related recruitment of the amygdala or IFG when testing hypothesis 2 (see Results), those two ROIs were also not pursued for hypothesis 3, to restrict number of analyses. Therefore, analysis of events of consistent (shared) positive and negative valence was restricted to our STS and precuneus ROIs *only*, to replicate previous literature (34-36). The STS AAL3 atlas ROI comprises the primary and frontal secondary auditory cortices, and the precuneus, for its role in narrative encoding (11).

## Neuroimaging data analysis
*Analysed in present study*

We conducted two separate neuroimaging analysis, i.e. we created two general linear model (GLM) design matrices in SPM.

### *Consistent (shared) emotion feature perception as conditions*

Each of the 8 sessions included had four conditions reflecting events with a high degree (>50%) of agreement among observers regarding emotion portrayal. These conditions were: high consistency positive valence (POS), high consistency negative valence (NEG), high consistency high arousal (arousal), and low agreement events (LAE) (i.e. <50% agreement for all emotion types). In addition to the 4 conditions, each session also included 6 regressors encoding volume-to-volume head motion (translations & rotations [x y z])).

This technique allowed us to estimate the degree to which the BOLD signal corresponded to a given explanatory variable, i.e. predicted BOLD response to each condition or motion-related signal change. We controlled for the autocorrelation by fitting a first order autoregressive (AR(1)) function for the fixed effects model. A high pass filter of 128Hz was applied to bypass scanner low frequency drift.

For all 8 sessions, POS, NEG and arousal were each compared to LAE, and equivalent reversed contrast. Finally, POS > NEG and reverse were also analysed. Importantly, the arousal condition was omitted from the 8$^{th}$ session as the arousal vector was colinear with the other predictors, meaning the variation in the signal was not being uniquely explained by the arousal predictor during the 8$^{th}$ session.

Each of these 1$^{st}$-level contrasts were taken forward to a random-effects group level model (i.e. a one sample t-test).

### *Musical features as parametric modulators for 3$^{rd}$ hypothesis*

The second GLM incorporated the musical features in the design matrix as parametric modulators. Here, for each of the same four conditions mentioned above, five 1$^{st}$ order (linear) parametric modulators were added. These parametric modulators (musical features) encoded trial-to-trial variation during the region's emotional perception brain activity on the basis of the strength of the musical feature's presence (i.e. they capture unexplained emotion perception-related signal variation that is linked to changes in the musical feature). Therefore, each of the four conditions are associated with 6 regressors (the initial predicted BOLD conditions, and five additional explanatory variables as parametric modulators), in addition to the 6 motion regressors, giving a total of 30 explanatory variables modelling the response to each session.

### *Significance in this study*

For hypothesis-led analysis, significance was defined at p<0.05 after family-wise error correction within small volumes defined by independent ROIs generated a priori. Therefore, we consider voxels significant where the difference in signal magnitude is above the threshold defined by gaussian Random Field Theory (gRFT), to ensure at most a 5% risk of the activity being false positives (see results tables).

For exploratory whole-brain analysis, results were considered significant if, using an uncorrected cluster-forming threshold, they survived whole-brain family-wise error on the basis of cluster extent (p<0.05 FWE). This means there is a 5% risk that our cluster corrected statistical t-map contains false positive clusters of the cluster threshold size (see results tables), as defined by gRFT. The peak level threshold here is more liberal; however, the false positive rate is controlled by the number of contiguous voxels, i.e. based on spatial extent of clusters.

# RESULTS

Table 3 below shows the demographic information of the participants of this study. Not all were musicians and some subjects had already seen the film before. Importantly, this table represents the participants before listening to the audio-description German version of Forrest Gump in a 7T fMRI scanner. That session was performed before the audio-visual Forrest Gump scanning used for this study.

**Table 3:** Participant (n=15) demographics

| Subject id | Gender | Age | Handed-ness | Hearing problems _current | Hearing problems _past | Music genre favourite | Musician | Seen Forrest Gump |
|---|---|---|---|---|---|---|---|---|
| 1 | m | 30-35 | r | n | n | Triphop | n | y |
| 2 | m | 30-35 | r | n | n | Bluegrass | n | y |
| 3 | f | 20-25 | r | n | y | Classical | n | n |
| 4 | f | 20-25 | r | n | n | Rock | y | n |
| 5 | m | 25-30 | r | n | n | Charts (Pop; Dance) | y | y |
| 6 | m | 20-25 | r | n | n | Rock | y | y |
| 9 | m | 30-35 | r | n | n | Metal | y | y |
| 10 | f | 20-25 | r | n | n | Alternative/Rock | n | y |
| 14 | f | 30-35 | r | n | y | Pop | n | y |
| 15 | m | 25-30 | r | n | n | n/a | n | y |
| 16 | m | 35-40 | r | n | n | Rock | n | n |
| 17 | m | 30-35 | r | n | n | Classic Rock | n | y |
| 18 | m | 30-35 | r | n | n | EBM/Industrial/Electro | n | y |
| 19 | f | 20-25 | r | n | n | Rock | n | y |
| 20 | f | 25-30 | r | n | y | Pop | y | y |

## I. PEARSON'S CORRELATION BETWEEN EMOTIONAL & MUSICAL FEATURES

To answer our first hypothesis, we tested on the whole research cut whether the fluctuating musical features such as timbre (spectral energy divided into three frequency bands: low, mid, high) and tempo (In beats per minute (BPM)) correlated significantly with the emotional portrayal agreement annotations, i.e. arousal and valence (see table 4). We principally found significant correlations between most features, the only feature that did not correlate significantly with most features is tempo – except positive and negative valence. Arousal shows negative correlation with positive valence, yet positive correlation with negative valence. Positive and negative valence demonstrate negative correlation.

Table 4: Correlation matrix between emotional collective perception and musical features during the research cut of Forrest Gump. * indicates p<0.05; ** indicates p<0.001786 (significance following Bonferroni correction)

|  | AROUSAL | POS VALENCE | NEG VALENCE | TEMPO | VOLUME | LOW FREQ | MID FREQ |
|---|---|---|---|---|---|---|---|
| POS_VALENCE | r=-0.2996 p<0.001** |  |  |  |  |  |  |
| NEG VALENCE | r=0.1823 p<0.001** | r=-0.2052 p<0.001** |  |  |  |  |  |
| TEMPO | r=-0.0101 p=0.3974 | r=0.0306 p=0.01** | r=-0.0463 p<0.001** |  |  |  |  |
| VOLUME | r=0.2582 p<0.001** | r=-0.1009 p<0.001** | r=0.0941 p<0.001** | r=-0.0084 p=0.4738 |  |  |  |
| LOW FREQ | r=0.3252 p<0.001** | r=-0.1218 p<0.001** | r=0.1104 p<0.001** | r= 0.0002 p=0.9883 | r=0.7525 p<0.001** |  |  |
| MID FREQ | r=0.3269 p<0.001** | r=-0.1406 p<0.001** | r=0.1988 p<0.001** | r=-0.0181 p=0.1234 | r=0.5842 p<0.001** | r=0.8194 p<0.001** |  |
| HIGH FREQ | r=0.2721 p<0.001** | r=-0.1082 p<0.001** | r=0.1675 p<0.001** | r=-0.0163 p=0.1668 | r=0.5171 p<0.001** | r=0.7283 p<0.001** | r=0.9074 p<0.001** |

## II. NEURAL ASSOCIATES OF AGREEMENT ON EMOTIONAL PERCEPTION WHILE MOVIE-WATCHING

### 1. Hypothesis-led analysis

#### 1.1. SUPERIOR TEMPORAL SULCUS ROI

Significant activity was observed in the hypothesis-driven superior temporal sulcus (STS) region of interest (ROI) under every condition of high inter-subject agreement on emotional perception, i.e. high agreement positive valence (POS) > low agreement events (LAE) (left STS: $pFWE_{Swc}$= 0.016, Z = 5.21, [-58, -22, 6]), NEG > LAE (left STS: $pFWE_{Swc}$= <0.001, Z = 5.28, [-60, -4, 2]), and HA arousal > LAE arousal (bilateral STS: $pFWE_{Swc}$= 0.021, Z = 5.16, [-44, -28, 12] & $pFWE_{Swc}$= 0.031, Z = 5.08, [-52, -2, 4]) & reversed contrast (right STS: $pFWE_{Swc}$= 0.019, Z = 5.18, [50, -18, -6]), as well as POS > NEG (bilateral STS: $pFWE_{Swc}$= 0.012, Z = 4.53, [56, -14, 8] & $pFWE_{Swc}$= 0.031, Z = 4.22, [-56, -20, 6]). Each contrast represents a weighted sum of the predictors (defined by the desired conditions summed to equal zero), divided by the standard deviance between the predictors.

#### 1.2. PRECUNEUS ROI

Furthermore, hypothesis-led analysis using the precuneus ROI demonstrated significant recruitment for all contrasts involving high agreement versus low agreement. We found evidence for bilateral precuneus activity for POS > LAE (pFWE$_{swc}$= 0.003, Z = 4.88, [-4, -64, 56] & pFWE$_{swc}$= 0.004, Z = 4.82, [12, -56, 16]), NEG > LAE ((pFWE$_{swc}$= 0.01, Z = 4.69, [-4, -60, 58] & pFWE$_{swc}$= 0.028, Z = 4.38, [14, -56, 16]), arousal > LAE (pFWE$_{swc}$= 0.005, Z = 5.41, [-10, -56, 56] & pFWE$_{swc}$= 0.001, Z = 5.75, [6, -64, 56]).

### 1.3. AMYGDALA & INFERIOR FRONTAL GYRUS ROIs

However, we found no evidence of significant activity for any condition within our two other ROIs, i.e. amygdala and inferior frontal gyrus. See table for further information.

## 2. Whole brain exploratory analysis
### 2.1. POSITIVE & NEGATIVE VALENCE RELATED BRAIN RESPONSES

We performed whole-brain exploratory analysis for all five conditions and obtained significant results for most contrasts. The occipital, parietal (including precuneus), temporal (including superior temporal gyrus) showed significant activity for POS > LAE (figure 1), NEG > LAE (see table 3 for significant clusters).

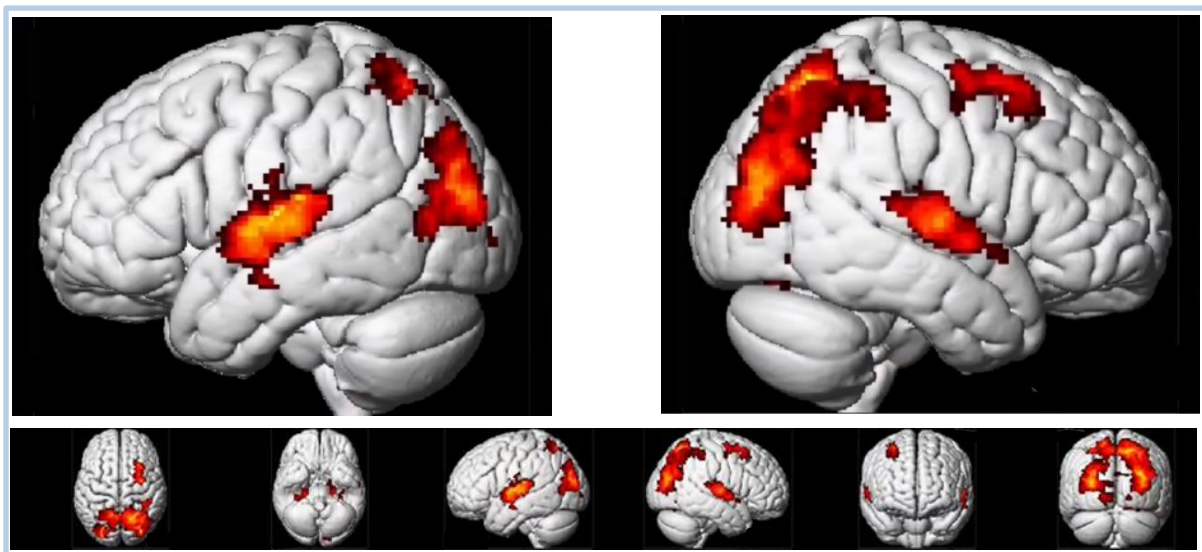

**Figure 5:** BOLD activity for high positive valence agreement (POS) versus low agreement (OTHER) cluster corrected for the whole research cut (see table 2 for coordinates).

**Table 5:** Brain regional activation for whole brain exploratory analysis after cluster correction, for POS > LAE & NEG > LAE (hypothesis 2). POS = high agreement on positive valence. NEG = high agreement on negative valence.

| Brain region labels | | POS > LAE | | |
|---|---|---|---|---|
| | | cluster | peak | |

|  |  | size | Z | x,y,z {mm} |
|---|---|---|---|---|
| BILATERAL OCCIPITAL LOBE | | 13804 | 5.54 | -30 -82 14 |
| | | | 5.45 | 34 -36 -14 |
| | | | 5.41 | 36 -80 20 |
| RIGHT TEMPORAL LOBE | | 1105 | 4.83 | 54 -16 6 |
| | | | 4.67 | 54 -2 -4 |
| | | | 4.21 | 38 -32 16 |
| RIGHT PARIETAL LOBE | | 988 | 4.63 | 30 -8 48 |
| | | | 4.59 | 26 0 56 |
| | | | 4.25 | 32 8 60 |
| | | **NEG > LAE** | | |
| | | cluster size | peak Z | x,y,z {mm} |
| LEFT TEMPORAL GYRUS | | 897 | 5.28 | -60 -4 2 |
| | | | 5.05 | -52 -10 -2 |
| | | | 4.48 | -60 -36 6 |
| RIGHT TEMPORAL GYRUS | | 244 | 4.97 | 34 -34 -14 |
| | | | 4.75 | 24 -42 -16 |
| | | | 4.11 | 28 -76 -8 |
| BILTAERAL OCCIPITAL LOBE, PRECUNEUS | | 4292 | 4.67 | 16 -60 28 |
| | | | 4.65 | -4 -72 58 |
| | | | 4.63 | 8 -46 56 |
| LEFT LIMBIC LOBE, PARAHIPPOCAMPAL GYRUS | | 202 | 4.65 | -28 -38 -6 |
| | | | 3.54 | -30 16 -16 |
| RIGHT FRONTAL LOBE, MIDDLE FRONTAL GYRUS | | 848 | 4.47 | 28 -2 46 |
| | | | 4.33 | 32 2 52 |
| | | | 4.32 | 24 -4 54 |
| RIGHT TEMPORAL GYRUS | | 309 | 4.31 | 54 -16 -4 |
| | | | 3.82 | 52 -12 4 |
| | | | 3.58 | 60 48 12 |
| RIGHT INFERIOR FRONTAL GYRUS | | 152 | 4.01 | 38 38 4 |
| | | | 3.89 | 34 56 6 |

### 2.2. AROUSAL & POS-NEG VALENCE RELATED BRAIN RESPONSES

Similarly, compared to LAE, HA arousal was associated with greater BOLD activity in the occipito-parieto-temporal lobe, as well as middle frontal gyrus, anterior cingulate and insula.

Finally, POS > NEG (figure 2) showed significantly greater occipito-temporal lobe activity, and important left frontal medial orbital recruitment.

Interestingly, NEG > POS (figure 3) revealed right middle frontal lobe activity.

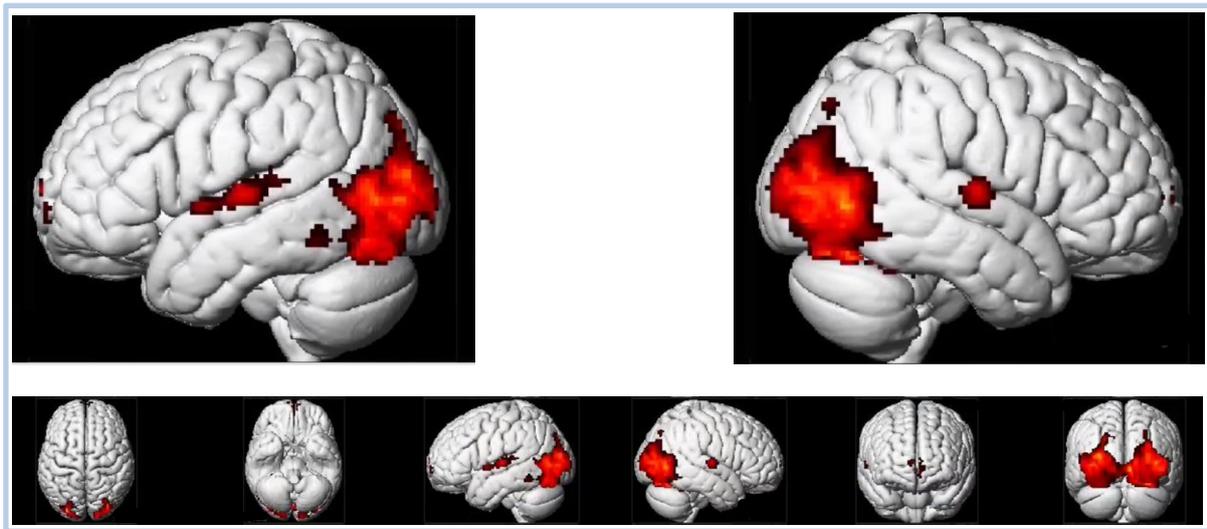

**Figure 6:** BOLD activity for high positive valence agreement (POS) versus high negative valence agreement (NEG) cluster corrected for the whole research cut (see table 2 for coordinates).

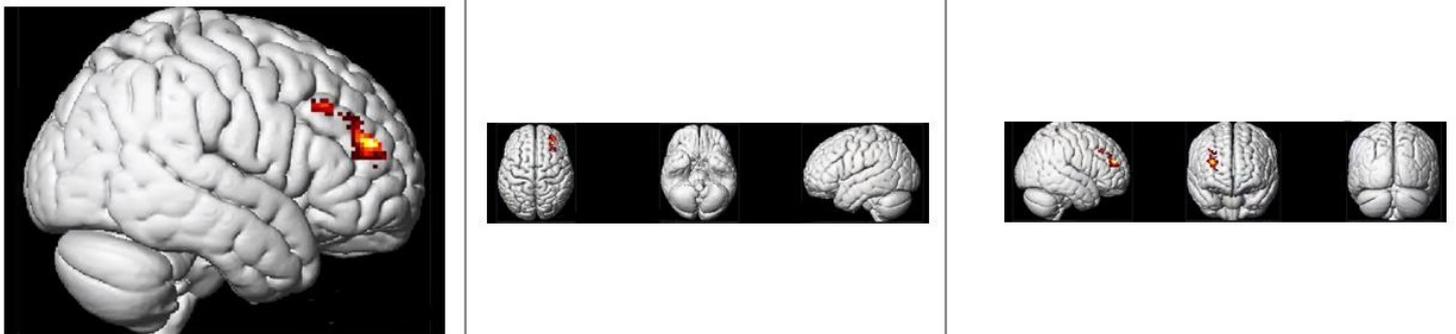

**Figure 7:** BOLD activity for high negative valence consistency versus high positive valence consistency cluster corrected for the whole research cut (see table 2 for coordinates).

**Table 6:** Brain regional activation for whole brain exploratory analysis after cluster correction, arousal > LAE, POS > NEG, NEG > POS (hypothesis 2). POS = high agreement on positive valence. NEG = high agreement on negative valence. STG = superior temporal gyrus. MTG = middle temporal gyrus.

|  |  | **AROUSAL > LAE** | | |
|---|---|---|---|---|
|  |  | cluster size | peak Z | x,y,z {mm} |

| Region | cluster size | peak Z | x,y,z {mm} |
|---|---|---|---|
| BILATERAL OCCIPITAL LOBE (PRECUNEUS, LINGUAL, PARAHIPPOCAMPAL GYRI) | 26365 | 6.42 | 26 -46 -12 |
| | | 5.99 | -30 -82 12 |
| | | 5.98 | 14 -86 24 |
| LEFT TEMPORAL LOBE (SUPERIOR TEMPORAL, POSTCENTRAL GYRI) | 2068 | 5.33 | -48 -16 10 |
| | | 5.16 | -44 -28 12 |
| | | 4.92 | -42 -16 18 |
| RIGHT FRONTAL LOBE (SUPERIOR & MIDDLE FRONTAL GYRI) | 1286 | 5.08 | 52 -2 -4 |
| | | 4.94 | 62 -18 12 |
| | | 4.89 | 40 -28 14 |
| RIGHT SUB-LOBAR | 148 | 4.93 | 0 16 -8 |
| LEFT MIDDLE FRONTAL GYRUS | 105 | 4.54 | -30 10 50 |
| LEFT SUB-LOBAR | 109 | 4.39 | -24 20 -2 |
| | | 3.93 | -30 22 6 |
| | | 3.75 | -36 14 12 |
| RIGHT TEMPORAL LOBE | 1173 | 4.36 | 24 12 54 |
| | | 4.35 | 44 -6 56 |
| | | 4.28 | 28 20 48 |
| LEFT ANTERIOR CINGULATE | 145 | 4.16 | 12 26 8 |
| | | 3.81 | -4 4 16 |
| | | 3.61 | 0 14 14 |
| RIGHT MIDDLE TEMPORAL GYRUS | 115 | 3.76 | 58 -48 -10 |
| | | 3.43 | 52 -34 -18 |
| | **POS > LAE** | | |
| | cluster size | peak Z | x,y,z {mm} |
| BILATERAL OCCIPITAL LOBE (LINGUAL, FUSIFORM GYRI, CUNEUS) | 10517 | 6.84 | 8 -86 -2 |
| | | 6.08 | 26 -80 -12 |
| | | 6.07 | 30 -70 -10 |
| RIGHT TEMPORAL LOBE(STG) | 178 | 4.53 | 56 -14 8 |
| LEFT TEMPORAL LOBE (STG) | 532 | 4.22 | -56 -20 6 |
| | | 3.99 | -44 -26 8 |
| | | 3.87 | -50 -2 2 |
| MIDDLE FRONTAL GYRUS | 137 | 4.03 | -2 66 -4 |
| | | 3.66 | 2 70 8 |
| | | 3.56 | -4 60 -12 |
| | **NEG > POS** | | |
| | cluster size | peak Z | x,y,z {mm} |
| RIGHT SUPERIOR, MIDDLE FRONTAL GYRUS | 326 | 4.93 | 26 28 34 |
| | | 4.5 | 32 46 22 |
| | | 3.79 | 26 36 32 |

## III. POSITIVE VALENCE & NEGATIVE VALENCE RELATED NEURAL RESPONSES ASSOCIATED WITH MUSICAL FEATURES (LOW FREQUENCIES & HIGH FREQUENCIES, RESPECTIVELY)

### 1. Hypothesis-led analysis

Significant associations were observed in the hypothesis-driven STS ROI for $NEG_{HF} > LAE_{HF}$, $LAE_{HF} > NEG_{HF}$ & $LAE_{LF} > POS_{LF}$ (see table 3). Significant music-related activity was also shown for the hypothesis-led precuneus ROI for $NEG_{HF} > LAE_{HF}$ & $POS_{LF} > LAE_{LF}$ (see table 4).

### 2. Whole brain exploratory analysis

#### 2.1 POSITIVE VALENCE RELATED BRAIN RESPONSE ASSOCIATED WITH LOW FREQUENCY

We performed whole-brain exploratory analysis for (figure 4 & 5) and found that the $POS_{LF} > LAE_{LF}$ contrast demonstrated significant recruitment of bilateral occipital lobe, extending into lingual gyrus, fusiform gyrus, precuneus, cuneus & BA19, as well as bilateral frontal lobe, including superior frontal gyrus. The $LAE_{LF} > POS_{LF}$ indicated significantly greater association low frequency energy. in bilateral temporal lobe, incorporating the superior temporal gyrus, also known as auditory cortex, as well as left inferior frontal gyrus (orbital) and right superior frontal gyrus (see table).

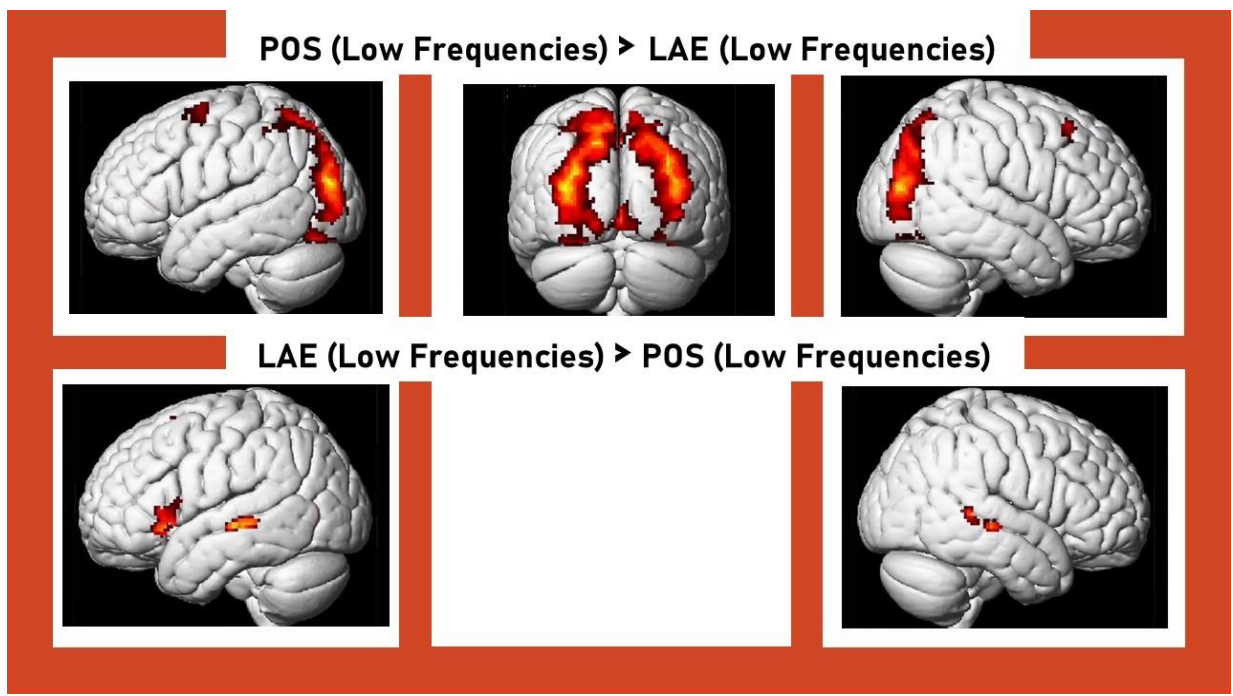

**Figure 8:** Cluster corrected significant activity modulated by low frequencies for high consistent (mostly shared) positive valence (POS) versus low consistent (less shared) onsets (LAE) and reverse during whole-brain exploratory analysis.

**Table 7:** Brain regional activation for whole-brain exploratory analysis after cluster correction for POS$_{LF}$ > LAE$_{LF}$ & LAE$_{LF}$ > POS$_{LF}$ (hypothesis 3; see appendix 1 & 2), with parametric modulators. POS = high agreement on positive valence. STG = superior temporal gyrus. MTG = middle temporal gyrus

| Brain region labels | | POS(LF) > LAE (LF) | | |
|---|---|---|---|---|
| | | cluster size | peak Z | x,y,z {mm} |
| BILATERAL OCCIPITO-TEMPORAL | | 11564 | 6.36 | -38 -86 18 |
| | | | 5.68 | 26 -42 -12 |
| | | | 5.52 | 38 -80 14 |
| RIGHT FRONTAL LOBE | | 101 | 4.57 | 26 22 54 |
| | | | 3.8 | 26 18 46 |
| LEFT FRONTAL LOBE (SUPERIOR & MIDDLE FRONTAL GYRI) | | 151 | 4.18 | -20 10 50 |
| | | | 4.05 | -22 -2 70 |
| | | | 3.65 | -26 4 66 |
| | | LAE (LF) > POS (LF) | | |
| | | cluster size | peak Z | x,y,z {mm} |
| LEFT INFERIOR FRONTAL GYRUS | | 181 | 5.02 | -48 22 -6 |
| | | | 3.78 | -56 18 4 |
| | | | 3.25 | -58 12 14 |
| LEFT FRONTAL LOBE (SUPERIOR FRONTAL GYRUS) | | 145 | 4.74 | -6 8 64 |
| | | | 3.93 | -2 2 58 |
| LEFT TEMPORAL LOBE (SUPERIOR & MIDDLE TEMPORAL GYRI) | | 164 | 4.39 | -56 -22 -4 |
| | | | 3.88 | -56 -32 0 |
| | | | 3.42 | -48 -34 -2 |
| RIGHT TEMPORAL LOBE (STG & MTG) | | 190 | 4.37 | 54 -24 -4 |
| | | | 4.05 | 52 -38 4 |

**2.2 NEGATIVE VALENCE RELATED BRAIN RESPONSE ASSOCIATED WITH HIGH FREQUENCY**

NEG$_{HF}$ > LAE$_{HF}$ recruited significant activity in the bilateral occipital lobe such as lingual gyrus, cuneus, precuneus, calcarine and pulvinar, as well as the temporal lobe, comprising superior, middle and inferior temporal gyri, bilateral thalami, and the corpus callosum. LAE$_{HF}$ > NEG$_{HF}$ showed significant activity in bilateral frontal lobe (involving superior frontal gyrus, middle frontal gyrus, inferior frontal gyrus and Brodmann Area 10), also known as frontal pole, as well as bilateral temporal lobe, which extends into the hippocampus, parahippocampal gyrus, fusiform gyrus, superior temporal gyrus, and middle temporal gyrus (see table).

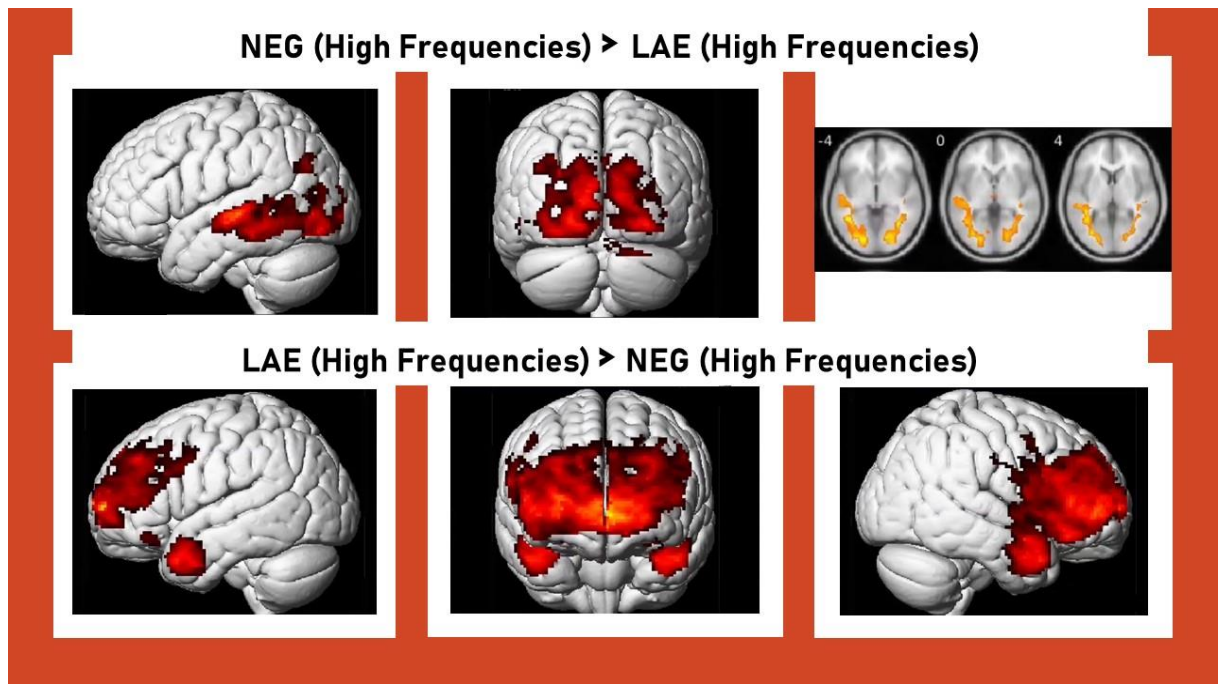

**Figure 9:** Cluster corrected significant activity modulated by high frequencies (HF) for high consistent (mostly shared) negative valence (NEG) versus low (less shared) consistent onsets (LAE) and reverse, during whole-brain exploratory analysis.

**Table 8:** Brain regional activation for whole-brain exploratory analysis after cluster correction for NEG$_{HF}$ > LAE$_{HF}$ & LAE$_{HF}$ > NEG$_{HF}$ (hypothesis 3; see appendix 3 & 4), with parametric modulators. NEG = high agreement on negative valence. STG = superior temporal gyrus.

|  | NEG (HF) > LAE (HF) | | |
|---|---|---|---|
|  | cluster size | peak Z | x,y,z {mm} |
|  | 19647 | 6.49 | -42 -48 -8 |
| BILATERAL OCCIPITO-TEMPORAL LOBE |  | 6.27 | -40 -42 8 |
|  |  | 6.07 | -30 -80 -10 |
|  | LAE (HF) > NEG (HF) | | |
|  | cluster size | peak Z | x,y,z {mm} |
|  | 19369 | 6.73 | -10 56 -4 |
| BILATERAL FRONTO-TEMPORAL LOBE |  | 6.54 | -4 60 0 |
|  |  | 6.3 | 30 56 12 |
| LEFT TEMPORAL LOBE (STG) | 1326 | 5.74 | -50 8 -28 |
|  |  | 5.56 | -40 16 -32 |
|  |  | 4.88 | -40 6 -36 |

# DISCUSSION

We sought to investigate three key points about emotional consensus while movie-watching.

1. Firstly, whether musical features (e.g. timbre and tempo) correlated with between-subject consistency regarding the arousal and valence of conveyed emotions.
2. Secondly, whether the emotional consensus correlated with brain activity within our ROIs, and more broadly in our exploratory whole-brain analysis.
3. Thirdly, whether a significant proportion of positive- and negative-valence related brain activity within ROIs (STS and precuneus) was associated with the spectral energy of music in the low frequency or high frequency ranges respectively.

We demonstrated that the STS and precuneus are important brain regions for audio-visual input and emotional encoding: remarkably associated with spectral energy in both low frequencies and high frequencies during positive and negative valence consensus, respectively. Importantly, our IFG ROI did not show significant activity, but frontal lobe gyri appeared during exploratory whole-brain analysis. Furthermore, frontal regional activity during movie-watching seems to relate to negative valence, and high arousal consensus. The amygdala did not show activity which could reflect our low sample size. Finally, we provided strong correlational evidence that the soundtrack of the film supports valence and high arousal events, and most likely by design.

## 1. EMOTIONAL FEATURES ARE SUPPORTED BY SOUNDTRACK TIMBRE

We found that musical features correlated significantly with the inter-subject agreement of emotional portrayal except for tempo which was not associated with consensus regarding emotional arousal. This suggests that the musical and sound timbre used in the movie has a crucial role in eliciting common perception of positive valence, negative valence, and arousal. Strong correlation between each frequency band and volume in RMS, and volume's strong correlation with emotional features is not surprising, as it underlies the evolving timbre of the film's soundtrack.

In truth, weak negative correlation between arousal and positive valence and a positive correlation between arousal and negative valence seems to suggest that the film induced more highly arousing events during moments of high negative valence consensus. Arousing and negative stimuli, like angry or fearful faces, tend to attract greater attentional resources, reducing subsequent attentional performance (56). This is increased in socially anxious individuals who risk negativity bias (13, 56). This recalls the attentional distribution theory which states that negative emotions reduce attentional scope, whereas positive emotions widen attentional breadth (14, 39, 40)

The association between emotional and acoustic features highlight an engineered over coincidental soundtrack. We must admit that the acoustic features extracted and analysed do not necessarily describe musical properties directly, as the study did not focus on scenes whose acoustic soundtrack were characterised by music. It is unclear whether increased activity reflects unexpectedness linked to increased neural activity, and thereby energy metabolism, or audio-visual emotional congruency that facilitates encoding (15). If soundtrack and emotions association represents emotional congruency, this intuition may allude to increased visual dominance during incongruency as seen with proven auditory cortex (STG) activity reduction, and bilateral fusiform gyrus recruitment (57).

However, semantic decisions (to support narrative) may have driven soundtrack design, while not prioritising sensory level emotional content (34, 36, 58, 59). To illustrate, the warfare soundtrack during the Vietnam scenes would be needed for storytelling purpose above the need to create suspense, although both elements may be inseparable during movie-watching.

A strong caveat from naturalistic stimuli is the poor experimental control risking reverse inference. Nonetheless, what is the meaning of the observed brain activity in this study?

## 2. NEUROIMAGING: STS MAY NOT PROCESS VALENCE

During actor audio-visual portrayal of emotions, STS was recruited for events where people commonly perceived positive valence, negative valence (including high frequency information), high arousal but also during low agreement moments (for high frequencies relative to POS and low frequencies relative to NEG).
Although relatively low amounts of acoustic features were extracted compared to those presented in the stimulus, focusing further investigations on one frequency band during NEG and POS, for example, might reveal a stronger comparative analysis. Nevertheless, our findings demonstrate the versatile activity of STS, believed to mirror other's emotions (e.g. actor's) (10), with strong evidence for audio-visual motion encoding (12, 43).
Unfortunately, we cannot tell whether those mechanisms occurred as we did not map visual information such as eye-gaze movement nor narrative or non-narrative objects to correlate to eye-gaze, or the visual rhythm of the edit. On the contrary, a simple, controlled task allows for confident assumptions from isolated regional activation, not possible if the condition is not met or the task is not performed well.

Strikingly, a surprising lateralization occurred for both negative and positive valence, compared to low agreement trials (left STS), which seems odd when positive valence recruits bilateral STS compared to negative valence. This may reflect the need for a greater sample size.

### 2.1. STS ENCODES HIGH AND LOW FREQUENCIES IRRESPECTIVE OF VALENCE

High frequency modulation of STS occurred regardless of valence consistency, therefore, may be unrelated to emotional congruency. However, due to STS recruitment during $NEG_{HF}$, and in line with previous literature, these results suggest an interesting avenue for future research. Prior studies showed congruent music enhanced suspense relative to incongruent or no music (45) and that fearful music was scream-like, and characterised as negative in valence and within high frequencies (60).
Analysing the rhythmic, pitch/melodic, harmonic as well as semantic information in the high frequency range (i.e. its content), in addition to measuring visual attention during NEG would reveal whether high frequency information aid reduction of attentional breadth (following attentional distribution theory (14, 39, 40)), thereby focusing viewers onto narrative content due to engineered emotional congruency. If no evidence, high frequencies might play a role outside the narrative, such as inducing ecologically valid impressions during scenes, e.g. with background sound effects. Dialogue has already been annotated by studyforrest researchers and may be a fruitful resource for this aim.

We did not find evidence of low frequency modulation of STS brain activity induced by the presence of high agreement of positive emotional stimuli (i.e. $POS_{LF}$). Importantly, low frequencies were integrated in the STS during low agreement events (LAE) compared to POS. The previous paragraph's suggestion can be extended to studying low frequency content, and during POS, as we are unsure in which frequency band was the narrative content mostly communicated.

These suggestions would also reveal to what extent valence, proven here to be associated with high or low frequency information, is linked to narrative content. Or even, during film-watching, what brain activity beyond valence defines aesthetic emotions, characterised by pleasantness bias, and emotional intensity regardless of low to high arousal spectrum (61).

## 2.2. PRECUNEUS ENCODES VALENCE FROM HIGH OR LOW FREQUENCY INFORMATION

The precuneus ROI findings demonstrate its role during POS (including low frequency modulation), NEG (including high frequency encoding) & high arousal processing. We can speculate that high or low frequencies transmit emotional information to the precuneus due to significant activity during negative and positive valence consensus, respectively. While precuneus processing of valence events is consistent with the literature (62), the findings mentioned do not reflect the role of musical features, narrative, locations displayed, or camera angles/movement used but suggest an interesting avenue for future research.

The precuneus is implicated in diverse visuospatial, episodic memory retrieval and perspective-taking functions e.g. storytelling-listening (48, 63). Musical episodic memory shows bilateral precuneus activation (64), although its function is not specific to music. Motor imagery such as imaginary finger tapping, or mental rotation of matrices (65), detection of high versus low pitch tones (66), i.e. visuospatial tasks, recruit bilateral precuneus (63).

Crucially, precuneus modulation due to high frequencies during NEG or low frequencies while POS requires additional data to be interpreted as the literature points.

## 2.3. FRONTAL REGIONS ACTIVE DURING AROUSAL AND NEGATIVE VALENCE

Strikingly, IFG ROI lacked significant activity for all conditions and contrasts. However, exploratory whole-brain analysis demonstrated IFG activity during NEG over LAE. The right frontal lobe (extending into superior and middle frontal gyri) was repeatedly shown active NEG for exploratory whole-brain analysis i.e. NEG > LAE & NEG > POS. HA high arousal showed similar frontal activity.

Significant correlation and similar brain activity between negative valence and arousal found in this study confirm previous findings that show negative stimuli tend to be more arousing (7).
These results may underlie a need for emotional regulation during negative valence. For instance, psychiatric illnesses such as Schizophrenia show a failure for cognitive change in the face of negative/unpleasant stimuli, possibly leading to increased negative emotionality (67). Principal Component Analysis estimated frontal areas as good predictors of emotional reappraisal, an essential feature for responding to negative stimuli (68).

Moreover, exploratory whole-brain analysis revealed left frontal gyri activation. Low agreement events compared to positive events demonstrated left inferior frontal gyrus - which includes the language processing region: Broca's Area (69) - as well as left superior and middle frontal gyri - strongly associated with Broca's language area (69, 70) - for POS relative to low agreement events. We can assume that the right handedness of our subjects corresponds to mainly left hemispheric language processing with low right hemispheric language processing (71). Therefore, voice-over narration, which does not occur only during consistent (shared) valence events, but probably also during low agreement events, would likely be processed in the left frontal gyri mentioned. This still requires further semantic analysis of the audio.

The aspect of the binary rating for arousal and valence and small sample observers may also explain the lack of significance in the amygdala ROI throughout all the emotional feature contrasts conducted. The amygdala encodes stimulus intensity irrespective of valence, although a preference for negative stimuli such as fear and disgust but also humour and sexual (i.e. high intensity emotions), and of gustatory-olfactory and visual modalities have been observed (72). The amygdala is not fully associated to arousal, as modest activity has been demonstrated with anger (which is highly arousing), yet social emotions, elicited by facial stimuli for example, are seen as strong predictors of amygdala activation (72). It is possible that studying the whole research cut removes the subtilies during social interactions of the film. Further investigation into specific scenes involving the actors is required to further amygdala's role in a naturalistic setting.

Exploratory whole-brain analysis displayed auditory cortex, precuneus, and fusiform gyrus activity for low frequencies during POS. Prominent recruitment of occipital lobe - containing fusiform gyrus, precuneus and cuneus — and auditory cortex do not reflect processes driving valence but rather the constantly changing rich visual information not controlled for in this experiment. For instance, fusiform gyrus encodes faces: it is damaged in prosopagnosia, i.e. impaired facial recognition (41, 42).
Unfortunately, not much research has uncovered the lingual gyrus' role in the occipital lobe, yet its activity has been associated with facial recognition (73), and mental imagery in secure attachment style individuals (relative to self-criticism management in avoidant attachment style) (74).

Finally, left anterior cingulate cortex (ACC; part of the Salience Network (75)) activity was observed during high arousal events. It is known that the region is linked to processing salient or behaviourally relevant cues (76), which are consistent with high arousal events. Interestingly, a previous study showed valence inference is ambiguous at climactic emotional intensity, however its arousal and intensity were consistently perceived. The study suggests that beyond a certain emotional intensity, emotional recognition is most ambiguous, but the saliency is at its highest, becoming akin to an alert signal (77). Emotional intensity measures should be tested in future naturalistic studies to understand the role of the ACC during high arousal events.

## CONCLUSION

Notwithstanding the various limitations highlighted throughout the discussion, in this thesis, we demonstrated an association between musical features and emotionally salient scenes and extended previous findings by showing that audio-visual valence and arousal are encoded in STS, and precuneus. Perceived valence seems to be associated with a number of spectral features, such as energy in low and high frequency range, yet the content that make up the frequencies still needs to be explored. Therefore, we were not able to make strong inferences about the musical content and scene-specific emotional perception fluctuation due to the exploratory nature of this study.

## LIMITATIONS AND FUTURE DIRECTIONS

Nevertheless, rich naturalistic stimuli offer a promising research avenue to explore interpersonal interactions and the role of music accompanying narrative. However, measurements of emotional perception must be improved. The consistent (shared) emotional perception annotations used here necessitate other emotion measurement tools. As mentioned, other elements not explored here, such as melody (27) and rhythm of the music (78, 79) and prosody (pitch changes irrespective of lexical content) (80) in terms of auditory information, and gestures and facial expressions (41, 42) in terms of visual input, also influence emotional communication and recognition.

For instance, using physiological measurements associated with emotions, such as heartbeat and breathing rates as conditions will further understand the role of the soundtrack in emotional perception. Furthermore, modelling the visual information is also essential in separating the impact of visual emotional communication from auditory emotional information. Hence, would scenes with higher ISC correlate with higher emotional consensus and stronger narrative understanding? The studyforrest team have eye-gaze as well as respiration and cardiac measurements which would nicely elucidate the basis of the findings of this thesis.

Problematically, the lack of strong experimental controls in current naturalistic studies result in an increased risk of lapsing into use of reverse inference. It is thus vital to explore narratives in shorter clips, as well as acquire as a fuller characterisation of inputs and associated cognitive processes recruited to reduce assumptions beyond measurements. Considerations for naturalistic stimuli design:

- Films with audio or visual impairment (reducing quality of perception, or simpler musical score or modified sound effects like reverberation, or a film cut without music)
- Removing one modality without modifying the other (avoid the audio-description use, as acoustic features are necessarily changed due to added narration)
- Matching emotional salience, basic characteristics in auditory and visual stream between scenes.

Individual difference has not been considered during this study. Several factors come to mind: ethnicity for cultural differences, anxiety trait (intolerance to uncertainty) for negative emotions bias (7), drug use for rewarding sensitivity e.g. heroin users are overly sensitive to heroin-themed films but have blunted reward reaction to erotic films (11).

Additionally, first episode psychosis patients demonstrated lower ISC compared to controls while viewing *Alice in Wonderland* (2010), potentially implying the importance of flexible emotional regulation during movie-watching (81), impaired in schizophrenia (67). Films may therefore at some point in the future prove useful as a psychiatric diagnostic tool, but could also be developed as therapeutic tool to train emotional regulation in the face of uncertainty and negativity. For example, does crying induced by audio-visual narratives improve emotional regulation?

Films like Forrest Gump convey 25 visual frames of information per second, whereas fMRI only acquires a full brain volume every 2 seconds. It is therefore essential to triangulate neuroimaging modalities, e.g. combining fMRI to the greater time resolution of electroencephalography (EEG) to approach analysis to human visual and auditory perception. A greater number of frequency bands would also approximate analysis closer to human ear frequency resolution (82). EEG recordings during naturalistic viewing are also needed to measure unexpectedness of incongruency shown to increase neural firing in the context emotionality agreement.

**SUMMARY**

This study has demonstrated that naturalistic stimuli experiments require stronger designs to use its research potential to induce emotions. This will unravel the potential therapeutic and diagnostic aspects of films. Valence was shown to be encoded by precuneus from audio information. Although STS activity showed some consistency with the literature, greater sample size and visual information mapping is required.

# ACKNOWLEDGEMENTS:

I would like to thank and honour Dr Owen O'Daly for his deeply passionate, pedagogical and highly informative approach to guiding this project. I learned a lot about scientific thinking and collaboration under his supervision. He has inspired me to continue doing science creatively and rigorously!

I would like to thank Mark Ghirardello for his tutoring and guidance for the thesis report.

I truly thank Leo Aimone for teaching me the basics of spectral extraction in Python and guiding me through it.

I am grateful for Leonardo Gavaudan's recommendations for Python programming.

Thank you to my parents for supporting me throughout this MSc.

# APPENDIX

*Note: All important supplementary information can be found in the Lab Book. Here, only the most salient information will be displayed.*

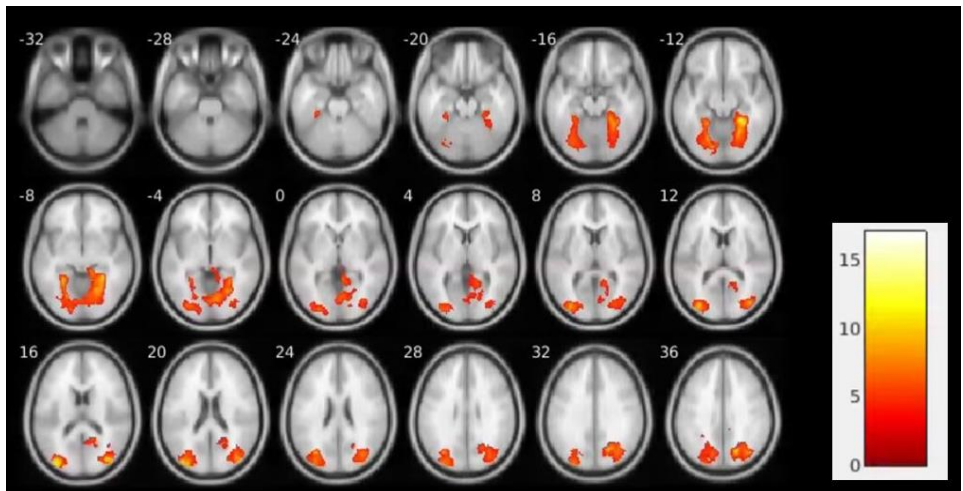

**Supplementary figure 1:** Brain regional activation for whole-brain exploratory analysis after cluster correction for $POS_{LF} > LAE_{LF}$. Coordinate values displayed in Table 7.

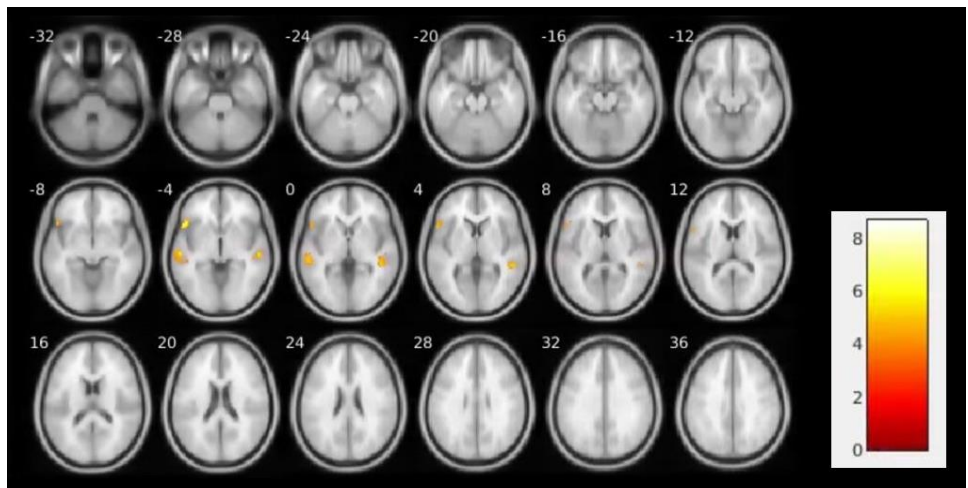

**Supplementary figure 2:** Brain regional activation for whole-brain exploratory analysis after cluster correction for $LAE_{LF} > POS_{LF}$. Coordinate values displayed in Table 7.

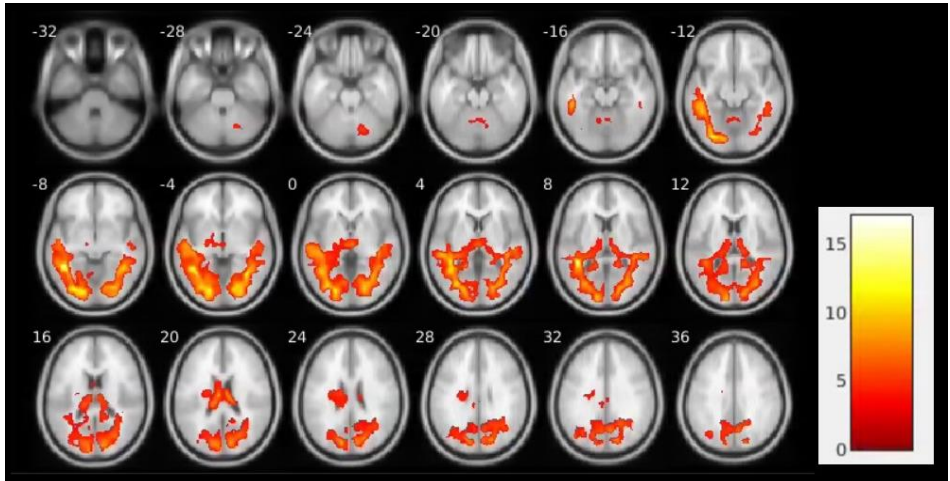
**Supplementary figure 3:** Brain regional activation for whole-brain exploratory analysis after cluster correction for NEG$_{HF}$ > LAE$_{HF}$. Coordinate values displayed in Table 8.

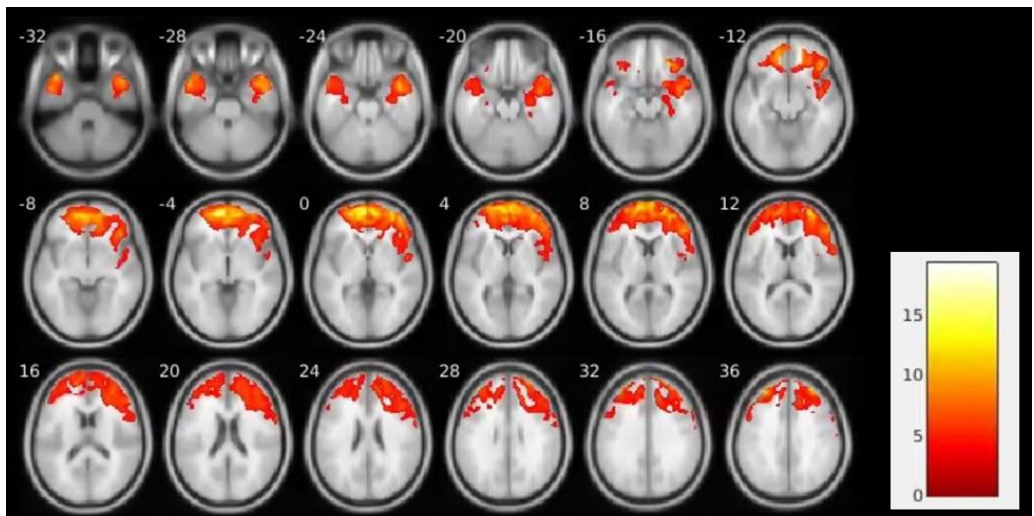
**Supplementary figure 4:** Brain regional activation for whole-brain exploratory analysis after cluster correction for LAE$_{HF}$ > NEG$_{HF}$. Coordinate values displayed in Table 8.